\documentclass[twocolumn]{webofc}

\usepackage[varg]{txfonts}   
\usepackage{hyperref}
\begin{document}
\title{HPRL – International cooperation to identify and monitor priority nuclear data needs for nuclear applications}

\author{
\firstname{E.}~\lastname{Dupont}\inst{1} \and
\firstname{M.}~\lastname{Bossant}\inst{2} \and
\firstname{R.}~\lastname{Capote}\inst{3} \and
\firstname{A.D.}~\lastname{Carlson}\inst{4} \and
\firstname{Y.}~\lastname{Danon}\inst{5} \and
\firstname{M.}~\lastname{Fleming}\inst{2} \and
\firstname{Z.}~\lastname{Ge}\inst{6} \and
\firstname{H.}~\lastname{Harada}\inst{7} \and
\firstname{O.}~\lastname{Iwamoto}\inst{7} \and
\firstname{N.}~\lastname{Iwamoto}\inst{7} \and
\firstname{A.}~\lastname{Kimura}\inst{7} \and
\firstname{A.J.}~\lastname{Koning}\inst{3} \and
\firstname{C.}~\lastname{Massimi}\inst{8} \and
\firstname{A.}~\lastname{Negret}\inst{9} \and
\firstname{G.}~\lastname{Noguere}\inst{10} \and
\firstname{A.}~\lastname{Plompen}\inst{11} \and
\firstname{V.}~\lastname{Pronyaev}\inst{12} \and
\firstname{G.}~\lastname{Rimpault}\inst{10} \and
\firstname{S.}~\lastname{Simakov}\inst{13} \and
\firstname{A.}~\lastname{Stankovskiy}\inst{14} \and
\firstname{W.}~\lastname{Sun}\inst{15} \and
\firstname{A.}~\lastname{Trkov}\inst{3} \and
\firstname{H.}~\lastname{Wu}\inst{6} \and
\firstname{K.}~\lastname{Yokoyama}\inst{7} \and
\lastname{the NEA WPEC Expert Group on the High Priority Request List for Nuclear Data}\inst{16}\fnsep\thanks{\email{wpec@oecd-nea.org}}
}

\institute{
CEA, Irfu, Universite Paris-Saclay, Gif-sur-Yvette, France \and
OECD, Nuclear Energy Agency (NEA), Boulogne-Billancourt, France \and
IAEA, NAPC Nuclear Data Section (NDS), Vienna, Austria \and
National Institute of Standards and Technology (NIST), Gaithersburg, MD, USA \and
Rensselaer Polytechnic Institute (RPI), Troy, NY, USA \and
CIAE, China Nuclear Data Center (CNDC), Beijing, China \and
Japan Atomic Energy Agency (JAEA), Tokai-mura, Japan \and
Istituto Nazionale di Fisica Nucleare (INFN) {\&} University of Bologna, Bologna, Italy \and
``Horia Hulubei'' National Institute for Physics and Nuclear Engineering (IFIN-HH), Bucharest - Magurele, Romania \and
CEA, Nuclear Energy Division (DEN), Cadarache, France \and
European Commission, Joint Research Centre, Geel, Belgium \and
Contractor, NAPC Nuclear Data Section (NDS), IAEA, Vienna, Austria \and
Karlsruhe Institute of Technology (KIT), Karlsruhe, Germany \and
SCK-CEN, Mol, Belgium \and
Institute of Applied Physics and Computational Mathematics (IAPCM), Beijing, China \and
www.oecd-nea.org/science/wpec/hprl
}

\abstract{
The OECD-NEA High Priority Request List (HPRL) is a point of reference to guide and stimulate the improvement of nuclear data for nuclear energy and other applications, and a tool to bridge the gap between data users and producers. The HPRL is application-driven and the requests are submitted by nuclear data users or representatives of the user's communities. A panel of international experts reviews and monitors the requests in the framework of an Expert Group mandated by the NEA Nuclear Science Committee Working Party on International Nuclear Data Evaluation Cooperation (WPEC). After approval, individual requests are classified to three categories: high priority requests, general requests, and special purpose requests (e.g., dosimetry, standards). The HPRL is hosted by the NEA in the form of a relational database publicly available on the web. This paper provides an overview of HPRL entries, status and outlook. Examples of requests successfully completed are given and new requests are described with emphasis on updated nuclear data needs in the fields of nuclear energy, neutron standards and dosimetry.
}
\maketitle
\section{Introduction}
The High Priority Request List (HPRL) for nuclear data has been established under the auspices of the Nuclear Energy Agency (NEA) in the 80s. The list is a compilation of the highest priority nuclear data requirements from nuclear data users. The purpose is to stimulate and guide measurement, nuclear theory and evaluation programmes. The rationale for the current list was established in the 2000s~\cite{Smith2005,Plompen2008} on the basis of more stringent criteria for adopting new requests. The list provides an international point of reference for both nuclear data users and producers, and its effectiveness in stimulating new measurements, evaluations and verification actions to meet the needs is well established.

\section{HPRL Organisation}
The HPRL is managed by an Expert Group mandated by the NEA Working Party on International Nuclear Data Evaluation Cooperation (WPEC) working in close collaboration with NEA for administrative items and for the maintenance of the HPRL database, tools and web pages.

\subsection{Governance}
A standing Expert Group is essential to maintain the HPRL as a point of reference in nuclear data research and development. The Expert Group consists of representatives from data evaluation project with expertise in the fields of nuclear data validation, evaluation and measurements. The Expert Group is responsible for managing the activities related to the HPRL, in particular for guaranteeing that the entries are up-to-date and well-motivated by current interests in the field of nuclear energy and other nuclear applications. The Expert Group is also responsible for stimulating follow-up to the entries and collecting the feedback provided by any of the related activities that may follow in response of a request.

\subsection{The Request Lists}
The HPRL is driven by nuclear applications on the basis of requests for specific nuclear data improvement made by users. All requests should be well justified in terms of impact on the application and of accuracy requirement with respect to the state-of-the-art experimental and evaluated data.

The HPRL consists of a list with high priority requests, a list with general requests and a list with special purpose requests divided in categories. Stringent criteria are applied for entries on the lists. These are evaluated by the Expert Group that makes the final decision for adopting a request in one of the lists.
\begin{itemize}
\item  A ``high priority” request is justified by sensitivity studies (or equivalent) and sufficiently documented.
\item  A ``general” request is well motivated for a specific quantity on a specific nucleus and is documented, but lacks a detailed backing by a sensitivity analysis or an impact study.
\item  A ``special purpose quantity” (SPQ) request in a well-defined category is of interest to a recognised important field of applied nuclear science for which it is essential to stimulate new activity.
\end{itemize}

The HPRL is subject to periodic review to monitor progress and assess the status of each entry. This is reflected by one of the following status:
\begin{itemize}
\item ``Work in progress'' for entries with ongoing experimental and theoretical activities.
\item ``Pending new evaluation or validation'' for entries that have already stimulated a lot of activities, but are not completed yet because of the lack of new evaluation or validation.
\item ``Completed'' (or ``Archived'') for entries that have been satisfied (or that are no longer relevant as priority need).
\end{itemize}

For the latter status the decision to terminate a request requires a consensus among the experts and care is taken not to affect too strongly the related activities.

\subsection{Database and Website}
The HPRL database, tools and web pages~\cite{HPRL} are hosted and maintained by the NEA. The technical implementation of the HPRL tools relies on Perl scripts connected to an Oracle relational database on the back and to HTML web pages on the front. They provide both a direct access to the request lists and to a search interface for querying the database, as well as to an online form for submitting a new request. The latter should contain various information on the requested improvement, the most important being the impact on the application, the accuracy requested on the nuclear data, and the justification with respect to the current state-of-the-art.

\section{HPRL Status}
The content of the database is summarized in table~\ref{content} with respect to categories and status. Although the ``Work in progress'' status dominates over the ``Completed'' entries, a sustained and continuous effort is currently dedicated for each of these entries. More information on the related activities is available in the HPRL web pages~\cite{HPRL}. The following subsections and tables give an overview of the nuclear data improvements requested.

\begin{table*}[htb]
\centering
\caption{Number of requests vs. categories and status. The values between parentheses are for individual requests in the SPQ category.}
\label{content}
\begin{tabular}{lcccc}
\hline
Status {\&} categories                         &  High priority  &  General  &  Special Purpose Quantity  &  Total      \\\hline
Work in progress                                 &       26            &      6         &          5 (64)                         &  37 (96)   \\
Pending new evaluation or validation &         4            &       0        &           0                                &    4          \\
Completed                                          &         4            &       4        &           0                                &    8          \\\hline
Total                                                    &       34            &     10        &           5 (64)                        &  49 (108) \\\hline
\end{tabular}
\end{table*}

\subsection{High priority and General requests}
The requests for fission and capture cross sections relevant to nuclear reactors are listed in table~\ref{nuclear}.

Requests for improvement of fission cross sections concern minor actinides (Np, Pu, Am, and Cm isotopes). With the exception of the request to improve the Np-237 fission, which is now completed, other requests are still work in progress after they have been put forward by WPEC Subgroup 26 (SG-26)~\cite{Salvatores2008}.

The requests for capture cross sections concern major actinides, such as the Big-3 (U-235,238 and Pu-239) and two other important fissile actinides (U-233, Pu-241), as well as two minor actinides (Pu-242, Am-241). These requests had also been put forward by SG-26. Additional recent capture requests concern structural material (minor isotopes of Cr~\cite{Koscheev2017}) and neutron absorbers (Gd-155,157~\cite{Rocchi2019} and Hf-nat). A significant achievement is the fulfilment of the requests for U-235 and U-238 capture cross sections, which were the subject of intensive collaborative works in the framework of the CIELO project~\cite{Chadwick2018,Capote2018}.

\begin{table*}[htb]
\centering
\caption{Fission and capture cross section requests relevant to nuclear reactors}
\label{nuclear}
\begin{minipage}[h]{0.99\textwidth}
\begin{tabular}{llll|llll}
\hline
Nuclide                                         & Reaction &  Half-life  &    Energy range     & Nuclide                    & Reaction                    &  Half-life          & Energy range\\\hline
Np-237$^\ast$                              & (n,f)        &  2.1 My    &   200 keV-20 MeV & U-233                      & (n,$\gamma$)            &  159 ky           &  Thermal-1 MeV \\
Pu-238                                          & (n,f)       &   88 y       &       9 keV-6 MeV    & U-235$^\ast$           & (n,$\gamma$)           &  704 My         &  100 eV-1 MeV    \\
Pu-240                                          & (n,f)       &   6.6 ky    &     0.5 keV-5 MeV   & U-238$^\ast$           & (n,$\gamma$)           &  $\sim$stable &  20 eV-25 keV     \\
Pu-241                                          & (n,f)       &   14 y       &   0.5 eV-1.35 MeV & Pu-239                     & (n,$\gamma$)           &  24 ky             &  1 meV-1.35 MeV \\
Pu-242                                          & (n,f)       &   375 ky   &   200 keV-20 MeV & Pu-241                     & (n,$\gamma$)           &  14 y              &  0.1 eV-1.35 MeV \\
Am-241                                         & (n,f)       &   432 y     &   180 keV-20 MeV & Pu-242                     & (n,$\gamma$),(n,tot) &  375 ky          &  0.5 eV-2 keV       \\
Am-242m                                      & (n,f)       &   141 y     &     0.5 keV-6 MeV  & Am-241                    & (n,$\gamma$),(n,tot) &  432 y            &  Thermal-Fast      \\
Cm-244                                         & (n,f)       &   18 y       &     65 keV-6 MeV   & Cr-50,53$^\dag$      & (n,$\gamma$)           &  stable           &  1 keV-100 keV     \\
Cm-245                                         & (n,f)       &   8.5 ky    &     0.5 keV-6 MeV  & Gd-155,157$^\dag$ & (n,$\gamma$),(n,tot) &  stable           &  Thermal-100 eV  \\
                                                      &              &                 &                               & Hf-nat$^\ast$           & (n,$\gamma$)           &  stable           &  0.5 eV-5 keV  \\\hline
\end{tabular}

\footnotesize{$^\ast$Request completed}
\footnotesize{$^\dag$Recent request ($\geq$2017)}
\end{minipage}
\end{table*}

The requests for other partial cross sections are compiled in table~\ref{other_xs}. Part of these requests had been submitted by SG-26, they concern inelastic scattering reaction on U-238, structural materials (Fe-56 and Si-28, which is now completed), and coolant of fast reactors (Na-23, Pb-206,207). There is also a recent request to improve the knowledge of the Bi-209 capture branching ratio (br) leading to the production of Po-210 in Pb-Bi eutectic under irradiation~\cite{Fiorito2018}. The remaining active requests are for K-39 activation and gas production cross sections in the NaK coolant of IFMIF-DONES~\cite{Simakov2017} and for needs related to neutron transport and criticality-safety (O-16, Pu-239~\cite{DeSaintJean2014}).

\begin{table}[htb]
\centering
\caption{Other partial cross section requests in various fields}
\label{other_xs}
\begin{minipage}[h]{0.99\textwidth}
\begin{tabular}{lll}
\hline
Nuclide                                               &  Reaction              &    Energy range        \\\hline
U-238                                                 &   (n,inl)                   &  65 keV-20 MeV      \\
Fe-56                                                 &   (n,inl)                   &  0.5 MeV-20 MeV    \\
Si-28$^\ast$                                       &   (n,inl)                   &  1.4 MeV-6 MeV      \\
Na-23                                                 &   (n,inl)                   &  0.5 MeV-1.3 MeV   \\
Pb-206,207                                        &   (n,inl)                   &  0.5 MeV-6 MeV      \\
Bi-209$^\dag$                                   &   (n,$\gamma$) br  &  500 eV-300 keV     \\
Si-28$^\ast$                                      &   (n,np)                   &  Threshold-20 MeV \\
K-39$^\dag$                                      &   (n,p),(n,np)          &  10 MeV-20 MeV     \\
Cr-52$^\ast$                                      &  (n,xd),(n,xt)          &  Threshold-65 MeV  \\
O-16                                                  &   (n,a)                    &  2 MeV-20 MeV       \\
Pu-239$^\dag$                                  &   (n,tot)                  &  1st resonance        \\
Au-197$^\ast$                                   &   (n,tot)                  &  5 keV-200 keV       \\\hline
\end{tabular}

\footnotesize{$^\ast$Request completed}
\footnotesize{$^\dag$Recent request ($\geq$2017)}
\end{minipage}
\end{table}

The HPRL is not limited to cross sections and table~\ref{other_quantities} shows requests for other quantities such as angular distributions, spectra and nubar. Improvement is requested for the H-2(n,el) energy-angle scattering probability distribution. There are also requests to improve the accuracy of important fission quantities, such as the Prompt Fission Gamma Spectra (PFGS) of the U-235 and Pu-239 major actinides, as well as the Prompt Fission Neutron Spectra (PFNS) of Am-243 and Cm-244 minor actinides, and also the nubar of U-233 and Pu-239,240.

\begin{table}[htb]
\centering
\caption{Requests for other various quantities}
\label{other_quantities}
\begin{minipage}[h]{0.99\textwidth}
\begin{tabular}{llll}
\hline
Nuclide              &  Reaction  &  Quantity                     &    Energy range    \\\hline
H-2                    &   (n,el)       &  $d/d\theta$                 &  0.1 MeV-1 MeV   \\
Fe-56$^\ast$     &   (n,xn)      &  DDX                           &  7 MeV-20 MeV    \\
U-235                &   (n,f)         &  $\gamma$ spectrum  &  Thermal-Fast       \\
Pu-239              &   (n,f)         &  $\gamma$ spectrum  &  Thermal-Fast       \\
Am-243             &   (n,f)         &  $n$ spectrum             &  Thermal-10 MeV  \\
Cm-244             &   (n,f)         &  $n$ spectrum             &  Thermal-10 MeV  \\
U-233                &   (n,f)         &  nubar                         &  Thermal-10 keV    \\
Pu-239$^\dag$ &   (n,f)         &  nubar                         &  Thermal-5 eV        \\
Pu-240              &   (n,f)         &  nubar                         &  200 keV-2 MeV     \\\hline
\end{tabular}

\footnotesize{$^\ast$Request completed}
\footnotesize{$^\dag$Recent request ($\geq$2017)}
\end{minipage}
\end{table}

\subsection{Requests for Special Purpose Quantities}
The category of Special Purpose Quantity was introduced in 2014 for requests individually lacking justifications to be accepted in the High priority or General categories, but with obvious generic value in a well-defined field. As of today, requests have been received in the fields of standards and dosimetry. A new subcategory dedicated to medical applications is foreseen.

\subsubsection{Standards}
In the SPQ subcategory dedicated to standards, requests have been received to further improve the n-p primary standard and to extend the fission standard at higher energy~\cite{Carlson2018} (see table~\ref{standards}).

\begin{table}[htb]
\centering
\caption{Requests for improvement of standards}
\label{standards}
\begin{minipage}[h]{0.99\textwidth}
\begin{tabular}{llll}
\hline
Nuclide                   &  Reaction  &  Quantity             &    Energy range   \\\hline
H-1                         &  (n,el)        &   xs, $d/d\theta$  &   10-20 MeV        \\
U-235,238$^\dag$ &  (n,f),(p,f)   &   xs                      &   100-500 MeV    \\\hline
\end{tabular}

\footnotesize{$^\dag$Recent request ($\geq$2017)}
\end{minipage}
\end{table}

\subsubsection{Dosimetry}
In the SPQ subcategory dedicated to dosimetry, recent requests have been adopted to address three different needs.

The first one is related to IRDFF validation~\cite{Trkov2013,Simakov2017dos}, which requires measurements of a number of spectrum-averaged cross sections (SACS) in well characterized Cf-252(sf) and/or U-235(n$_{th}$,f) neutron fields (see table~\ref{dos-irdff}).

\begin{table*}[htb]
\centering
\caption{Recent requests for spectrum-averaged cross sections relevant for the improvement and validation of IRDFF}
\label{dos-irdff}
\begin{minipage}[h]{1.0\textwidth}
\begin{tabular}{llc|llc|llc}
\hline
Nuclide     & Reaction          &  Spectra$^\ddag$ & Nuclide   & Reaction           &  Spectra$^\ddag$  & Nuclide    & Reaction           &  Spectra$^\ddag$  \\\hline
Co-59       & (n,$\gamma$)  &       Cf                   &  Rh-103  &  (n,inl)Rh-103m &      U5                     & Al-27        &  (n,2n)               & Cf {\&} U5          \\
Th-232     & (n,f)                  &       Cf                   &  Tm-169  &  (n,2n)                &      U5                     & Ti-46        &  (n,2n)              & Cf {\&} U5          \\
Zn-67       & (n,p)                 &       Cf                   &  Cu-65    &  (n,2n)                &      U5                     &  Sn-117    &  (n,inl)Sn-117m & Cf {\&} U5         \\
Mo-92      & (n,p)Nb-92m    &       Cf                    &  Mn-55   &  (n,2n)                &      U5                     &  Ti-47       &  (n,np)               & Cf {\&} U5          \\
Ni-60        & (n,p)                &       Cf                    &  Ni-58     &  (n,2n)                &      U5                     &  Ti-48       &  (n,np)               & Cf {\&} U5          \\
Fe-54       & (n,a)                &       Cf                    &  Am-241 &  (n,f)                   & Cf {\&} U5               &  Ti-49       &  (n,np)               & Cf {\&} U5          \\
As-75       & (n,2n)              &       Cf                    &  P-31      &  (n,p)                  & Cf {\&} U5               &  Fe-54      &  (n,2n)               & Cf {\&} U5          \\
Y-89         & (n,2n)              &       Cf                    &  U-238    &  (n,2n)                & Cf {\&} U5               &  Tm-169   &  (n,3n)               & Cf {\&} U5          \\
Mn-55      & (n,$\gamma$) &       U5                   &  In-115    &  (n,2n)In-114m   & Cf {\&} U5               &  Bi-209     &  (n,3n)               & Cf {\&} U5          \\
U-238       & (n,$\gamma$) &       U5                  &  Pr-141   &  (n,2n)                & Cf {\&} U5               &  Co-59      &  (n,3n)               & Cf {\&} U5          \\
Cu-63       & (n,$\gamma$) &       U5                  & Cr-52      &  (n,2n)                & Cf {\&} U5               &                 &                           &                           \\
La-139     & (n,$\gamma$) &       U5                  &  Na-23     &  (n,2n)                & Cf {\&} U5               &                 &                           &                           \\\hline
\end{tabular}

\footnotesize{$^\ddag$Cf and U5 stand for PFNS of Cf-252(sf) and U-235(n$_{th}$,f), respectively}
\end{minipage}
\end{table*}

The second one is for improvement of the high-energy tail of the Pu-239(n$_{th}$,f) PFNS that requires measurements of spectrum-averaged cross sections in a well characterised Pu-239 PFNS field~\cite{Capote2016}. The SACS should be measured for well-known high-threshold dosimetry (n,2n) reactions on F-19, Mn-55, Co-59, Ni-58, As-75, Y-89, Zr-90, Nb-93, I-127 and Tm-169.

The third and last one concerns the dosimetry of neutron sources that requires the measurement of both low-threshold, Sn-117(n,inl), and high-threshold cross sections with plateaus located between $\approx$15~MeV and $\approx$150~MeV~\cite{Trkov2013,Simakov2017dos} (see table~\ref{dos-he}).

\begin{table*}[htb]
\centering
\caption{Recent requests for improvement of cross sections relevant to the dosimetry of neutron sources}
\label{dos-he}
\begin{tabular}{lll|lll}
\hline
Nuclide        &  Reaction             &    Energy range (MeV)      & Nuclide      &  Reaction         &    Energy range (MeV)       \\\hline
Sn-117         &   (n,inl)Sn-117m  &             5 - 10                     & Cu-63        &  (n,2n)              &           20 - 100                   \\
Fe-nat          &   (n,x)Mn-54        &           15 - 100                   & Fe-54        &  (n,2n)              &           15 - 100                   \\
Y-89             &   (n,p)                  &           15 - 100                   & Au-197      &  (n,xn) x=3-5    &           20/Threshold - 100   \\
Y-89             &   (n,xn) x=2-4      &           15/Threshold - 100  & Tm-169     &  (n,xn) x=2-3    &           15 - 100                    \\
Ti-nat           &   (n,x)Sc-46         &           15 - 100                   & Bi-209       &  (n,xn) x=3-10  &           20/Threshold - 150   \\
Ti-nat           &   (n,x)Sc-47         &           15 - 100                   & Co-59        &  (n,xn) x=3-5    &           20/Threshold - 150   \\
Ti-nat           &   (n,x)Sc-48         &           15 - 100                   & Rh-103      &  (n,xn) x=4-8    &           Threshold - 150        \\
Lu-175         &   (n,xn) x=2-4      &           15/Threshold - 100  & La-139       &  (n,xn) x=4-10 &           Threshold - 150         \\
Nb-93          &   (n,xn) x=2-4      &           15/Threshold - 100  &                   &                         &                                             \\\hline
\end{tabular}
\end{table*}

\section{Conclusion}
In order to efficiently stimulate and guide nuclear data improvement the HPRL follows a twofold strategy. First of all, the HPRL aims to be a reference tool in support to experimental, theoretical and evaluation projects aiming at improving nuclear data. In parallel, the HPRL aims to bridge the gap between nuclear data users and nuclear data producers (evaluators and experimentalists).

The role of the NEA WPEC Expert Group in close collaboration with the nuclear data community is to ensure that the HPRL is up-to-date and properly reflecting the priority needs in the field of nuclear energy and other nuclear applications, as well as advances aiming at answering those needs.

Nuclear data users are invited to contribute with new requests either by email or using the online form~\cite{HPRL}. Feedback on HPRL entries from both users and producers of nuclear data is obviously welcome.


\end{document}